\documentclass{PoS}
\usepackage{natbib}
\usepackage{amsmath}

\title{Statistical analysis of archival Vela~X-1 data}

\ShortTitle{Statistical analysis of Vela~X-1}

 \author{{Felix F\"urst},$^a$ Ingo Kreykenbohm,$^a$ J\"orn Wilms,$^a$ Peter Kretschmar,$^b$ 
 Dmitry Klochkov,$^c$ Andrea Santangelo,$^c$ and R\"udiger Staubert\,$^c$\\
  \llap{$^a$}Dr. Karl-Remeis-Sternwarte Bamberg, ECAP, Universit\"at Erlangen-N\"urnberg, Sternwartstr.~7, 96049~Bamberg, Germany\\
 \llap{$^b$}European Space Agency, European Space Astronomy Centre, Villafranca del Castillo, P.O.~Box~78, 28691~Villanueva de la Ca\~{n}ada, Madrid, Spain\\
 \llap{$^c$}Kepler Center for Astro and Particle Physics, Institut f\"ur Astronomie und Astrophysik, Universit\"at T\"ubingen, Sand~1, 72076~T\"ubingen, Germany\\
E-Mail: \email{felix.fuerst@sternwarte.uni-erlangen.de } }

\abstract{We present a detailed analysis of all archival INTEGRAL data of the accreting X-ray pulsar Vela~X-1. We extracted lightcurves in several energy bands from 20\,keV up to 60\,keV. The lightcurves show that the source was found in very active as well as quiet states. During the active states several giant flares were detected. For these states spectra between 5\,keV and 120\,keV were obtained. The spectra of the active states were found to be significantly softer than those from the quiet states. We performed a statistical analysis of the flaring behavior. The resulting log-normal distribution of the intensity of Vela X-1 shows that the source spends most of the time at an average flux level of 300\,mCrab but also that the distribution extends well up to more than 2.0\,Crab. 
}

\FullConference{7th INTEGRAL Workshop\\
		 September 8-11 2008\\
		 Copenhagen, Denmark}

\begin{document}

\section{Lightcurve and Hardnessratio}
\begin{figure}
   \centering
   \includegraphics[width=0.85\textwidth]{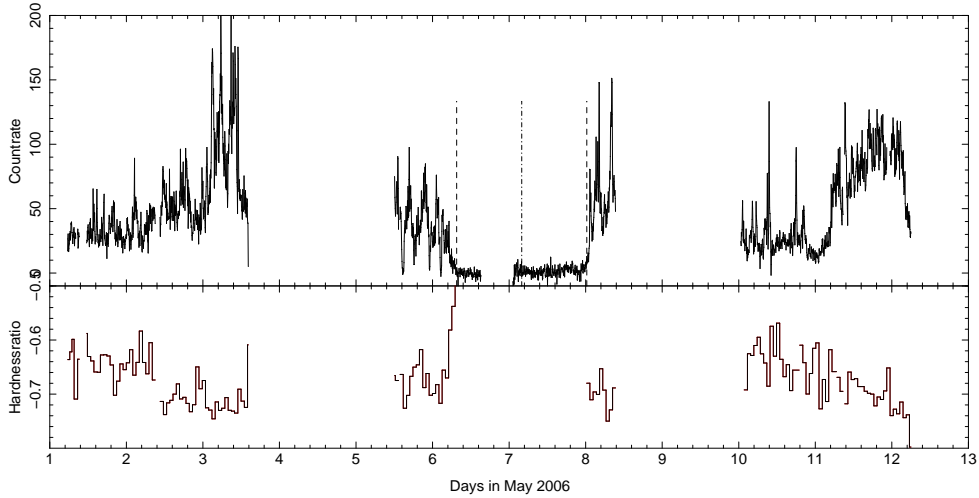}
   \caption[Lightcurve 2006 May]{Lightcurve in the 20--60\,keV band and hardness-ratio for data from Revs. 433--440, i.e. from 2006 May 02 to 2006 May 24. For the hardness-ratio the hard band from 40--60\,keV and the soft band from 20--30\,keV was used.}
   \label{fig:lc_20-60_4xx.pdf}
 \end{figure}

In this work, the accreting HMXB Vela~X-1 is studied, which has a pulse period of $\sim$283.5\,sec and an orbital period of $\sim$9.0\,days. Overall $\sim 3.6$\,Msec of data were analyzed, taken mostly by ISGRI aboard \textit{INTEGRAL}, which is a large amount of data compared to other measurements. Parts of the data have been published already \citep[see][]{kreykenbohm08a, schanne07a}, presenting lightcurves and spectra. A repetition of these works was not intended, instead a more statistical ansatz was taken. In that approach events were not regarded individually, but all events were binned together in some kind of common grid. Lightcurves for the statistical analysis were extracted from ISGRI in three different energy bands: 20--40\,keV, 40--60\,keV, and 20--60\,keV. The time-resolution was chosen to be 283.5\,sec to average each
data-point over one pulse period in order to eliminate these fluctuations. Fig. \ref{fig:lc_20-60_4xx.pdf} shows the lightcurve of the 20--60\,keV band for data from the revolutions 433--440 as an example for the analyzed data. The vertical dashed lines show the beginning and the end of the eclipse according to the ephemeris data from \cite{kreykenbohm08a}, and the dash-dotted line shows the
respective center of eclipse.  The lightcurve was extracted using \texttt{ii\_light}.

\begin{figure}
   \centering
   \includegraphics[width=0.8\textwidth]{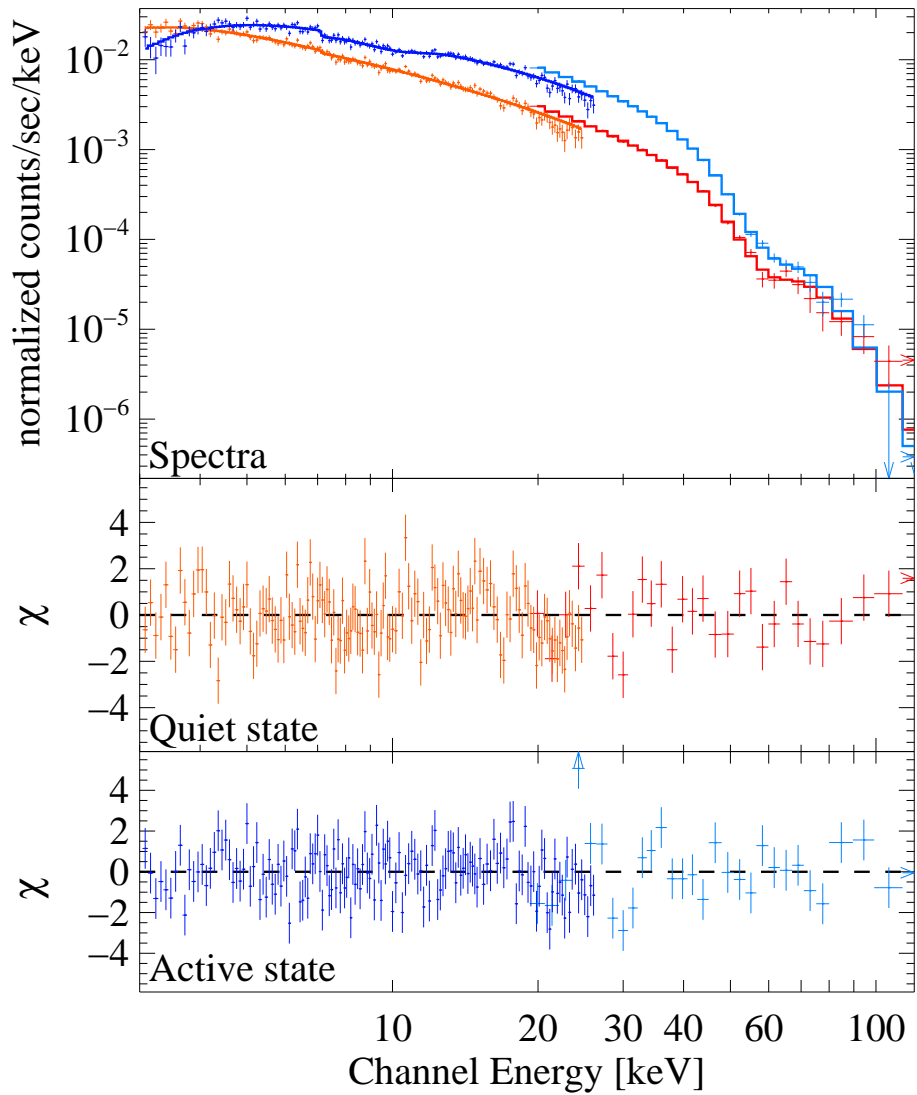}
   \caption[\textit{INTEGRAL} Vela~X-1 spectrum]{Spectra combined of JEM-X and ISGRI data. The blue spectrum shows the active state, the red one the quiet state. The second and third panel are the according residuals.}
   \label{fig:spectracomb}
 \end{figure}

The second panel of Fig. \ref{fig:lc_20-60_4xx.pdf} shows the hardness-ratio for the respective bins. The hardness-ratio is given on the timescale of Science Windows (ScWs) to smooth out the noise. The hardness-ratio $\vartheta$ is calculated as
\begin{equation}
 \vartheta = \frac{H-S}{H+S}
\end{equation}
where $H$ is the countrate in the hard band between 40--60\,keV and $S$ is the countrate in the soft band between 20--40\,keV. A larger value of $\vartheta$, i.e., a less negative value, is called harder, where as smaller values of $\vartheta$ are called softer. The hardness-ratio is a gauge for the physical processes taking place and is indicating the shape of the spectra. In Fig.~\ref{fig:lc_20-60_4xx.pdf} it is evident that in the bright states Vela~X-1 becomes distinctly softer. The two main parts of interest are at  May 01 -- May 3.5  and May 10 -- May 12. We note, that in the bright part of the second time range, from May 11 -- May 12, Vela~X-1 does not show distinct flaring, but is overall almost three times as bright as in first part, from May 10 -- May 11. To analyze this behavior more closely, we took a look at the spectra in these time ranges. As the exposure times are relatively small, we combined the spectra from both active states of Fig.~\ref{fig:lc_20-60_4xx.pdf}  and the spectra from both quiet states. The active state is shown in blue, while the quiet state is shown in red. The spectra were fitted in XSPEC \citep{xspec} independently but using the same model, namely
\begin{equation}
I(E) = C \cdot ( I_{\text{phabs}}  \cdot I_{\text{cont}} ) \cdot I^1_{\text{gauabs}} \cdot I^2_{\text{gauabs}} 
\end{equation} 
This model describes a photoabsorbed continuum, superposed by two Gaussian
absorption lines, one to model the CRSF at $\sim$58.6\,keV, the other to model a
yet unexplained feature at $\sim$10\,keV. The constant $C$ is needed, as
spectra from two different instruments, namely JEM-X and ISGRI were
used. The calibration between these instruments is inconsistent, so
that slightly different absolute fluxes are measured. All other
parameters of the model spectrum are the same for both
instruments. The soft part of the spectra was extracted using JEM-X from 3.0\,keV to 26.5\,keV, the harder part emerges from ISGRI data between 19.0\,keV and 130.0\,keV.
The second and third
panel of Fig. \ref{fig:spectracomb} show the respective residuals as
reduced $\chi^2$ value.  As shown in Fig.~\ref{fig:spectracomb} the photo index of
the quiet state is about 1.5 times as high as the one of the active
state. Following \cite {becker05a} this is due to the fact that less photons were Comptonized to high energies in
the accretion column of the neutron star. This is the case, when more material is accreted, e.g. when the absorption coefficient rises, as the photons are either emitted or absorbed before they can be scattered to high energies. A larger accretion rate additionally leads to an increased brightness in X-rays. This theory can consequently explain the effect that the spectrum
softens while Vela~X-1 is getting brighter. Another hint for an increased absorption coefficient is the fact that in the active
state, the $N_\mathrm{H}$ value of the photoabsorption part of the model is higher, which is a
direct effect of more material in the vicinity of the neutron
star. The energy and the shape of the CRSF does not change within the
uncertainties, so that an influence of the magnetic field on the shape
of the spectrum can be ruled out.

\section{Flaring behavior}
\label{sec:velaflares}
\begin{figure}
   \centering
   \includegraphics[width=0.8\textwidth]{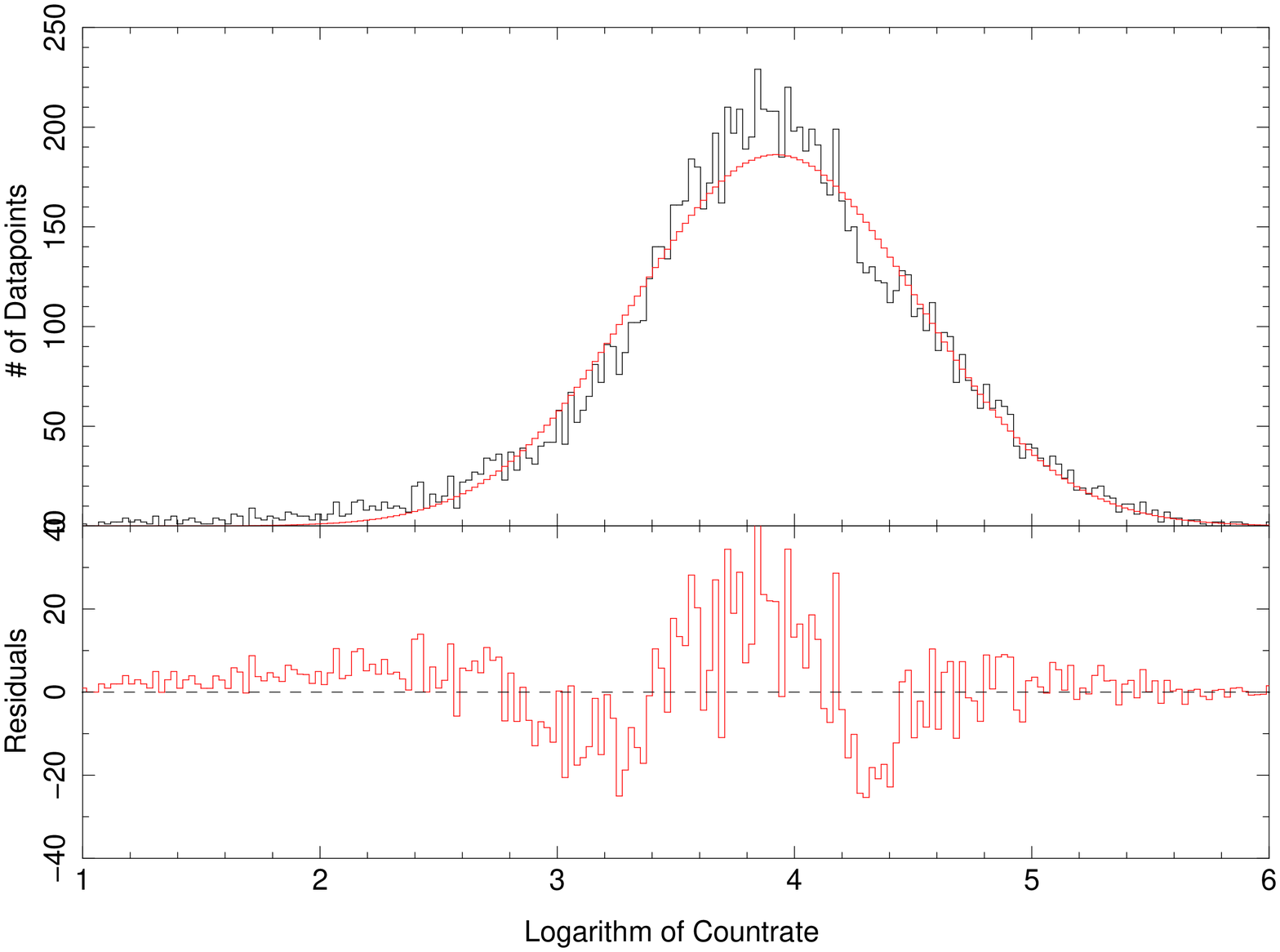}
   \caption[Histogram of LC in 20--60\,keV band, 1 Gaussian]{Histogram of the lightcurve of the 20--60\,keV band, binned to 256 bins. The red curve shows the best fit single Gaussian.}
   \label{fig:lcratehistall.ps}
 \end{figure}

During their observeration of Vela~X-1 \cite{kreykenbohm08a} could measure giant flares in the lightcurve of Vela~X-1, which are much brighter than the usual flares of the object. To understand if such giant flares are explainable by the same physical process as the other, common flares in Vela X-1, we analyzed the flaring behavior for the whole data statistically. The flares of Vela~X-1 have typical durations of some hours. To get a good coverage of these flares, the chosen temporal resolution of the lightcurve of 283.5\,sec is well suited. The lightcurve data were binned logarithmically into 256 countrate bins after eliminating all datapoints measured during the eclipse of Vela~X-1. Fig. \ref{fig:lcratehistall.ps} shows the histogram for the whole 20--60\,keV band. The solid line is a Gaussian fitted to the data, representing a log-normal distribution. 

\begin{figure}
   \centering
   \includegraphics[width=0.8\textwidth]{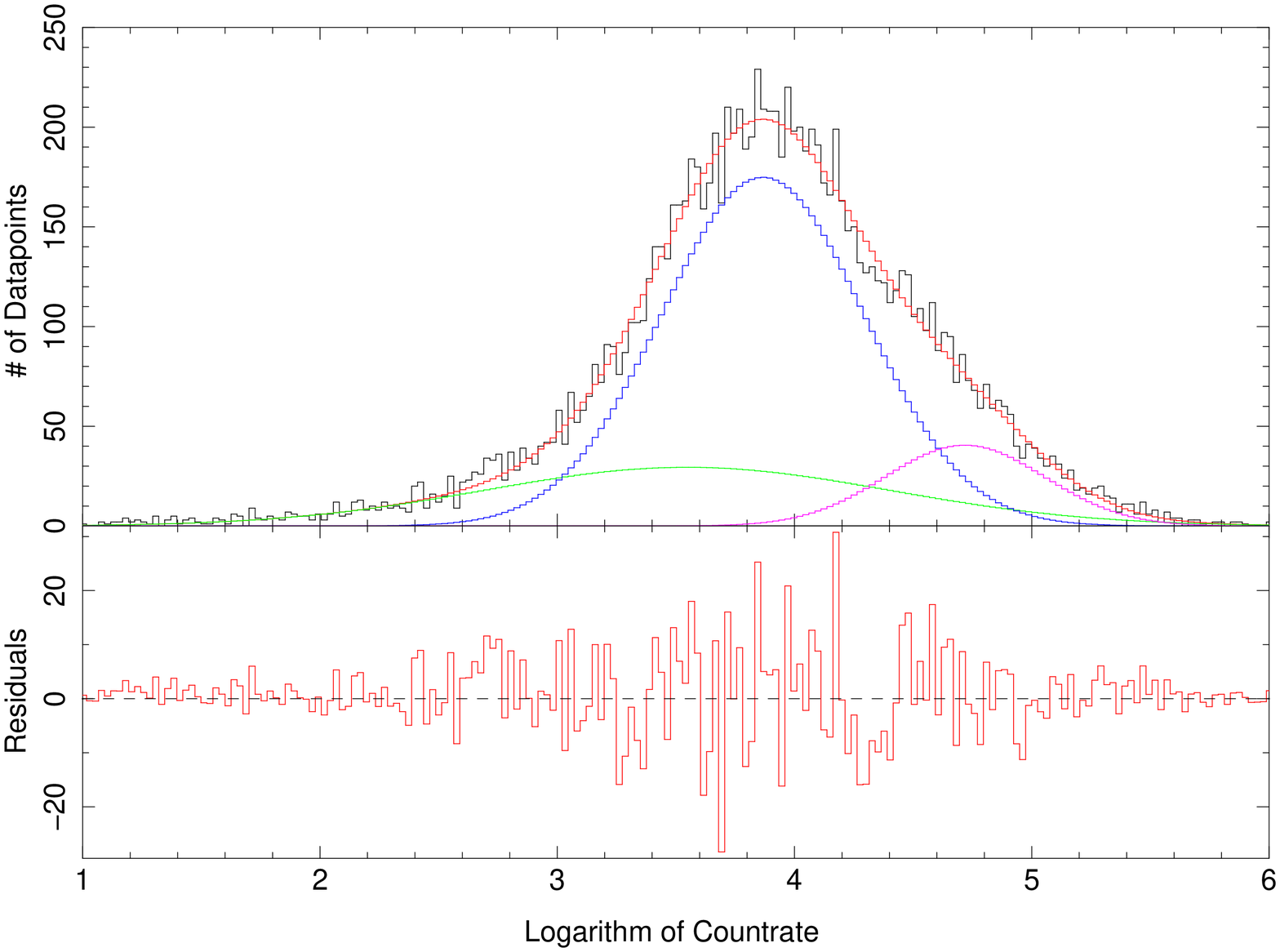}
   \caption[Histogram of LC in 20--60\,keV band, 3 Gaussians]{Histogram of the lightcurve of the 20--60\,keV band, binned to 256 bins. The red curve shows the best fit consting of 3 Gaussians. The blue curve shows the Gaussian \#1, the violet one Gaussian \#2 and the green one Gaussian \#3.}
   \label{fig:lcratehistall3g.ps}
 \end{figure}

Fig. \ref{fig:lcratehistall.ps} shows that the flaring follows closely a log-normal distribution. Even though the system is quite different, it is interesting to note that \cite{uttley01a} for short time-scales and \cite{poutanen08a} for long time-scales found a similar log-normal distribution when analyzing the flaring behavior of Cygnus X-1. These results were confirmed by \cite{gleissner04a}. A log-normal distribution can be achieved by multiplying $N$ random subprocesses. The emerging distribution will be log-normal distributed \citep{uttley05a}.
Even if the log-normal distribution is a rather good description of the histogram, small offsets at especially high and low countrates are apparent in Fig. \ref{fig:lcratehistall.ps}. These offset mean that the process is not produced by completely multiplicative components, but small additive components are a not-negligible. To improve the fit, two additional Gaussians where added to the model.  This improves the fit drastically, as shown in Fig. \ref{fig:lcratehistall3g.ps}. The probability to measure extremely bright flares with more than 300\,counts\,sec$^{-1}$ in the 20--60\,keV band is with one Gaussian $\sim$0.10\%, while when using 3 Gaussians it is $\sim$0.16\%. This is a rather small number but far from impossible, so that we can conclude that bright flares are rare but not once-in-a-lifetime events.

\begin{figure}
   \centering
   \includegraphics[width=0.8\textwidth]{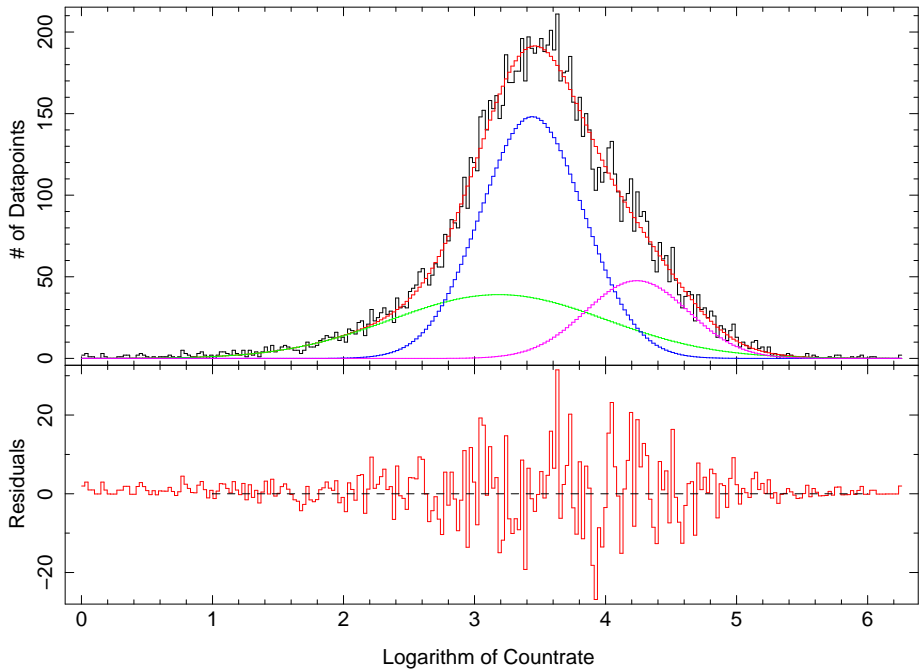}
   \caption[Histogram of LC in 20--30\,keV band]{Histogram of the lightcurve of the 20--30\,keV band, binned to 256 bins. The red curve shows the best fit consting of 3 Gaussians. The blue curve shows the Gaussian \#1, the violet one Gaussian \#2 and the green one Gaussian \#3. }
   \label{fig:lcratehistsoft3g.ps}
 \end{figure}

\begin{figure}
   \centering
   \includegraphics[width=0.8\textwidth]{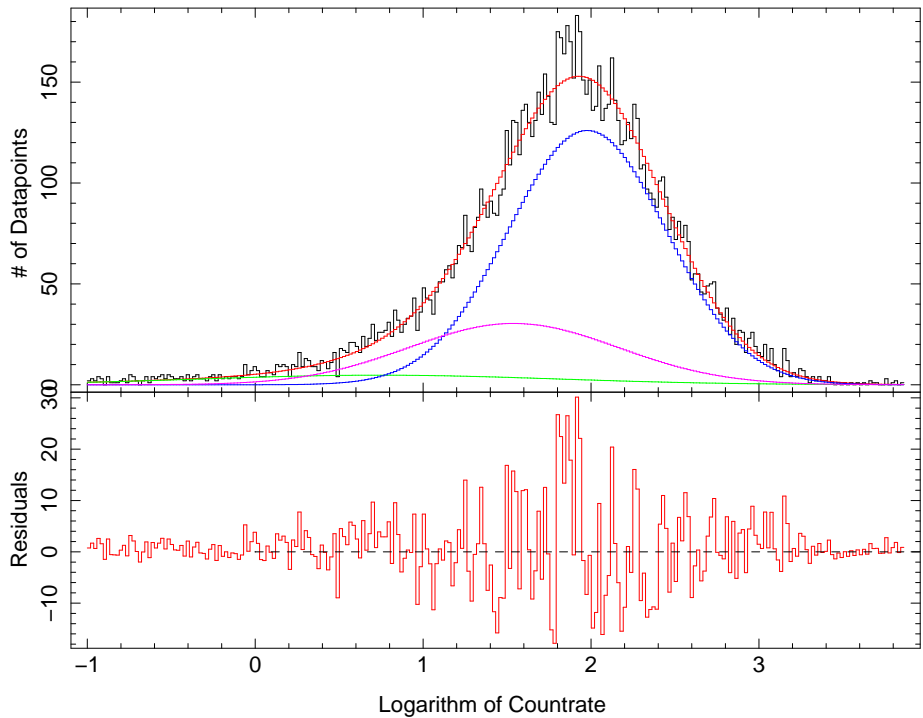}
   \caption[Histogram of LC in 40--60\,keV band]{Histogram of the lightcurve of the 40--60\,keV band, binned to 256 bins. The red curve shows the best fit consting of 3 Gaussians. The blue curve shows the Gaussian \#1, the violet one Gaussian \#2 and the green one Gaussian \#3. }
   \label{fig:lcratehisthard3g.ps}
 \end{figure}

A fit with three independent Gaussians was also done for the logarithmic binned lightcurves of the soft 20--30\,keV band and the hard 40--60\,keV band, see figs. \ref{fig:lcratehistsoft3g.ps} and \ref{fig:lcratehisthard3g.ps}. Gaussian \#2, describing the brighter part of the distribution is even more pronounced in the data of the soft band than it is in the data for the overall band. On the other hand, no Gaussian is  needed to model the active end of the histogram for the hard band. Even more, the main Gaussian \#1 models the distribution almost perfectly, so that the two other become very small. Gaussian \#3, responsible for the quiet end continuum is almost constant throughout all energy bands, only its overall strength is smaller in the hard band. When comparing the peak values of the Gaussians the power-law-like spectrum of Vela~X-1 must be regarded, which means that the hard band has intrinsically lower countrates than the soft band. Assuming a factor of Soft / Hard of about 5.6 leads to a value of $\log(5.6) = 1.73$ which must be added to the center value of the Gaussians of the hard band. This approximation would lead to a value of $\sim 3.7$ for the main Gaussian of the soft band, which is near the actual measured value. 

\begin{figure}
   \centering
   \includegraphics[width=0.8\textwidth]{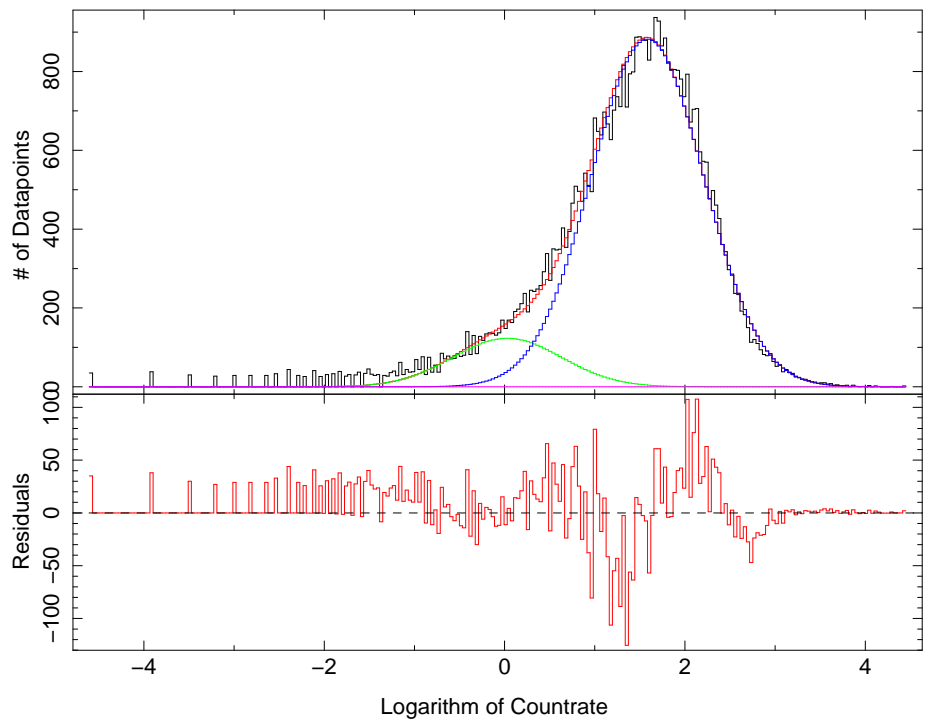}
   \caption[Histogram of LC of the ASM in the 2--10\,keV band]{Histogram of the lightcurve of the ASM of \textit{RXTE} in the 2--10\,keV band, binned to 256 bins. The red curve shows the best fit consting of 3 Gaussians. The blue curve shows the Gaussian \#1, the violet one Gaussian \#2 and the green one Gaussian \#3. Gaussian \# 2 as vanished completely and the bright flank is perfectly modelled with only one Gaussian. }
   \label{fig:ctratehisto_asm.eps}
 \end{figure}

To compare our findings to more measurements data from the All-Sky Monitor (ASM) aboard the ``Rossi X-Ray Timing Explorer'' \textit{RXTE} was analyzed. The ASM performs 90\,min dwells on a irregular basis onto Vela~X-1, providing a good statistic in the 2--10\,keV energy band. We performed the same analysis of the ASM data as of the ISGRI data, with binning the logarithm of the countrate to 256 bins after removing datapoints in the eclipse. A model of three individual Gaussians was then fitted to the data. The 2--10\,keV range is clearly softer than the soft band of ISGRI so that an even stronger bright excess would be expected. Fig.~\ref{fig:ctratehisto_asm.eps} shows, however, that no bright excess is measured. Even more the second Gaussian, responsible for modelling the bright excess in the \textit{INTEGRAL} data has vanished completely. One possible explanation for this behavior would be that the photo-absorption in the active state is drastically increased. As the cross-section of the photo-absorption is declining fast with increasing energy, X-ray photons of about 20\,keV are only slightly absorbed, while the 2--10\,keV photons are strongly absorbed. The increased absorption is evident in the spectrum shown in Fig. \ref{fig:spectracomb}. This increased absorption means that even very bright flares can not be seen in the soft band, because the soft X-rays can not penetrate the surrounding medium. A higher absorption in the active states can easily be explained by the fact that more material is accreted onto the neutron star, which means that it moves through denser parts of the stellar wind.  Another reason that no bright excess is apparent in the histogram of the ASM might be the inadequate sampling rate of the ASM. As shown by \cite{staubert04a} flares which were clearly visible in \textit{INTEGRAL} data do not show up in the ASM data. Even more, the ASM erroneously detects fluxes on the order of the flaring fluxes while Vela~X-1 was in eclipse \citep{kreykenbohm08a}. This reduces the trust which should be put into the histogram of the ASM data further. 

We already noted that brighter parts of the lightcurve are softer. This agrees very nicely with the observation that the second Gaussian is vanishing in the bright part of the histogram for the hard band, while it is undoubtedly clear in the soft band. This result could mean that two different physical processes are at work when producing the X-rays of Vela~X-1. One process might be responsible for the most of the variability, having the same strength in all energy bands. This variability could be due to small variations in the accretion stream, without changing the overall configuration of the system. The other process just produces soft X-rays, but with distinctly higher countrates. This process is not so common as the first one and might be explained by extremely increased accretion which than increases the optical depth in the accretion column dramatically. As explained above, this would lead to a softer spectrum. Much more work on this topic is necessary to understand the physical processes.

\end{document}